\documentclass[lettersize,journal]{IEEEtran}
\usepackage{amsmath,amsfonts}
\usepackage{algorithmic}
\usepackage{algorithm}
\usepackage{array}
\usepackage[caption=false,font=normalsize,labelfont=sf,textfont=sf]{subfig}
\usepackage{textcomp}
\usepackage{stfloats}
\usepackage{url}
\usepackage{verbatim}
\usepackage{graphicx}
\usepackage{cite}
\begin{document}

\title{NLOS Dies Twice: Challenges and Solutions of V2X for Cooperative Perception}

\author{~\IEEEmembership{Lantao Li and Chen Sun, Senior Member, IEEE}
\thanks{Lantao Li and Chen Sun are with Beijing Laboratory, R\&D Center China, Sony (China) Limited, 100027, Beijing, China. (Email: lantao.li@sony.com; Chen.Sun@sony.com )}
}

\markboth{SUBMISSION TO IEEE Vehicular Technology Magazine}%
{Shell \MakeLowercase{\textit{et al.}}: A Sample Article Using IEEEtran.cls for IEEE Journals}

\maketitle

\begin{abstract}
Multi-agent multi-lidar sensor fusion between connected vehicles for cooperative perception has recently been recognized as the best technique for minimizing the blind zone of individual vehicular perception systems and further enhancing the overall safety of autonomous driving systems. This technique relies heavily on the reliability and availability of vehicle-to-everything (V2X) communication. In practical sensor fusion application scenarios, the non-line-of-sight (NLOS) issue causes blind zones for not only the perception system but also V2X direct communication. To counteract underlying communication issues, we introduce an abstract perception matrix matching method for quick sensor fusion matching procedures and mobility-height hybrid relay determination procedures, proactively improving the efficiency and performance of V2X communication to serve the upper layer application fusion requirements. To demonstrate the effectiveness of our solution, we design a new simulation framework to consider autonomous driving, sensor fusion and V2X communication in general, paving the way for end-to-end performance evaluation and further solution derivation.
\end{abstract}


\section{Introduction}
\IEEEPARstart{A}{utonomous} driving, a highly anticipated feature of future vehicles, has been pursued and publicized by numerous manufacturers, tier 1 suppliers and research facilities. Despite the significant progress made recent years in related fields, such as data-driven learning-based driving techniques \cite{ref1,ref2} or computer vision tasks \cite{ref3,ref4,ref5}, challenges exist still before reliable full autonomous vehicles come to a reality. Especially for the vehicle perception system, a single-agent perception system is inherently susceptible to occlusions, putting the vehicle at risk for irreparable losses at any time, although tests demonstrate an extremely low probability of accidents. Static objects such as vegetation, buildings, and road constructions, combined with vehicles themselves, pose non-line-of-sight (NLOS) issues for almost all types of sensors (e.g., lidar, camera, radar).

Recent studies explored the benefits of sensor data sharing from multiple viewpoints of neighboring vehicles \cite{ref6,ref7,ref8}, exploiting the advent of V2X technologies to augment the actual detection area of individual vehicles, minimizing the NLOS issues for perception systems. However, these works \cite{ref6,ref9} all assumed the quality of service (QoS) provided by V2X communication techniques as a set of constant values (i.e., data rate, delay budget, packet error rate) based on the road test statistics, without serious consideration on how the environment might dynamically influence the wireless communication condition. In practical V2X road test activities, the impact of obstacles on the V2X communication is no less significant than their impact on the cooperative perception, meaning the NLOS issue again applies to V2X wireless channels \cite{ref10,ref11,ref12} and should be taken into serious consideration, as depicted in Fig. 1.  

In this paper, we bridge cooperative perception and V2X communication, proposing a novel approach for conducting sensor fusion in a dynamic wireless communication environment. To efficiently initiate the essential sensor fusion process between mobile vehicle actors, we introduce the abstract perception matrix matching (APMM) method with low computation and communication costs. To counteract the NLOS influence, we apply an optimized layer-agnostic relay node selection policy to V2X-enabled vehicles while considering both mobility and height factors. To examine how the environment might influence cooperative perception and wireless communication as a system for autonomous driving, we also develop a new co-simulation framework for solution derivation, performance evaluation and result analysis. To the best of our knowledge, this paper is the first to illuminate the overall design of vehicular networks for cooperative perception under dynamic wireless conditions and the unique cross-platform comprehensive simulation framework that involves communication, mobility and sensing. In summary, the main contributions are as follows:
\begin{itemize}
\item{We identify the problem of NLOS for both vehicle-mounted sensor(s) and V2X wireless communication.}
\item{We introduce the APMM, which can be implemented in various V2X messages and trigger sensor fusion/cooperative perception service based on quantifiable necessity.}
\item{We present our optimized mobility height determination (MoHeD) method on a relay node selection policy for sensor fusion, improving the reliability and continuity of the sensor fusion service.}
\item{We propose a new simulation framework to conduct and evaluate the overall performance of V2X services, connecting physical mobility factors, wireless communication environment, sensor properties and autonomous driving policy more closely and more realistically.}
\end{itemize}

\begin{figure}[!t]
\centering
\includegraphics[width=3.4in]{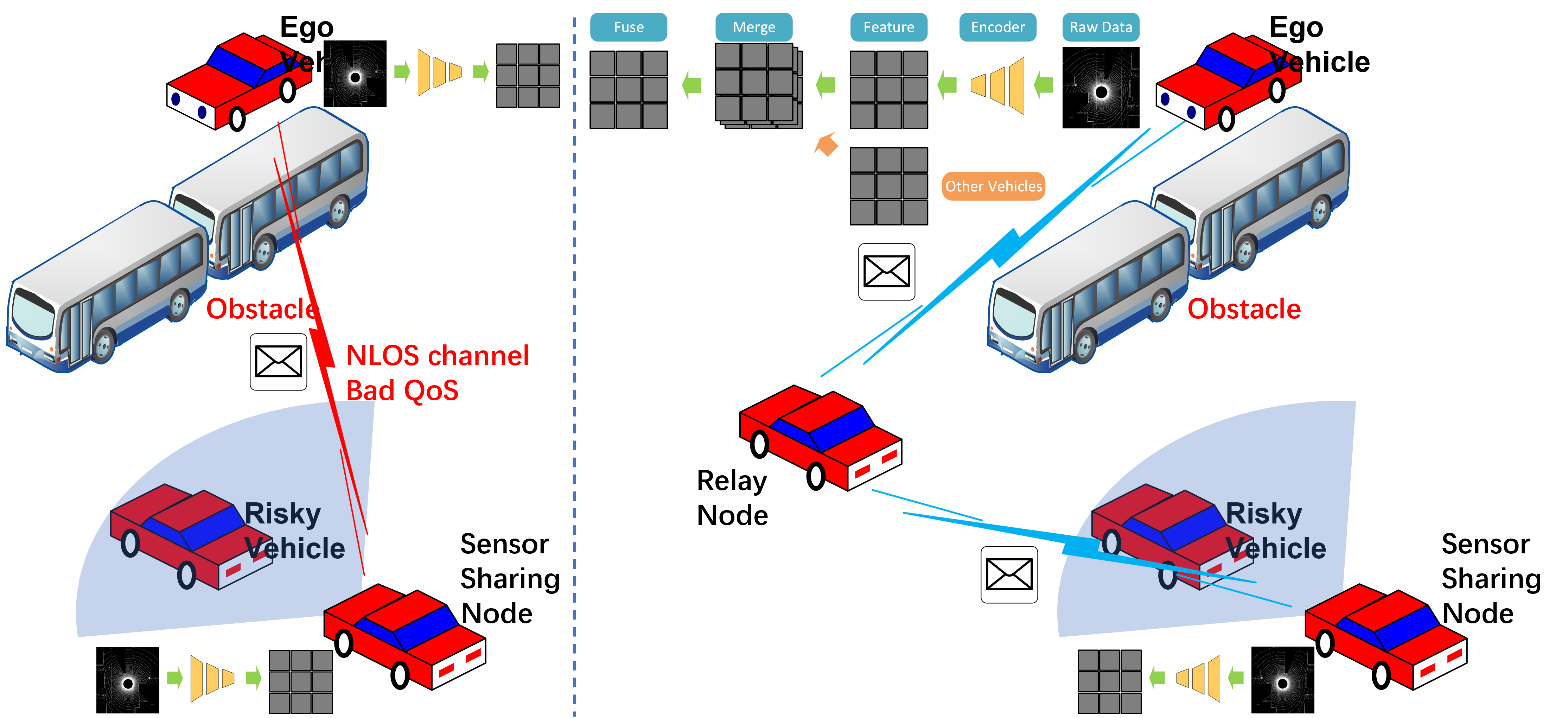}
\caption{The motivation to tackle the NLOS issue during sensor fusion data transmission, as the direct link in the left half cannot support data transmission for sensor fusion, while the right half could by exploiting nearby relay nodes to enhance perception.}
\label{fig_1}
\end{figure}

\section{Related works}
{\bf{Cooperative Perception for Autonomous Driving:}} Imitation learning \cite{ref32,ref33} has been adopted as the part of the baseline paradigm for autonomous driving after pioneered by Pomerleau \cite{ref13}, cooperative perception \cite{ref14,ref15} realized by sharing the raw or processed sensor data among vehicle actors minimizes the size of perception systems’ blind zone area, recent fusion transformer\cite{ref7,ref16} has brought the state-of-the-art performance to the next level, and COOPERNAUT \cite{ref9} combines the essence of both fields and provides a more general end-to-end solution.

{\bf{Connected Vehicles:}} V2X D2D, the most prominent technique in the Vehicular Ad-Hoc Network (VANET) domain\cite{ref17}, is expected to leverage the potential of sensor sharing, cooperative awareness and other critical applications by the industry \cite{ref20, ref21}, and the frequent NLOS transmission hinders reliability of the V2X technique massive deployment. Studies \cite{ref18,ref19} concentrating on different layers’ design of D2D technology, especially the relay selection mechanism, have also been carried out to mitigate the shadow of obstacles.

\section{Preliminary}
The overall system model can be divided into three sub-modules: mobility, wireless and perception. Mobility data in the designed scenario serve as the physical foundation for both the wireless and perception sub-modules, revealing whether the planned collision is avoided or not. Cooperative perception relies on wireless technology to acquire sufficient raw or processed sensing information as input to the autonomous driving agent.

\begin{figure}[!t]
\centering
\includegraphics[width=3in]{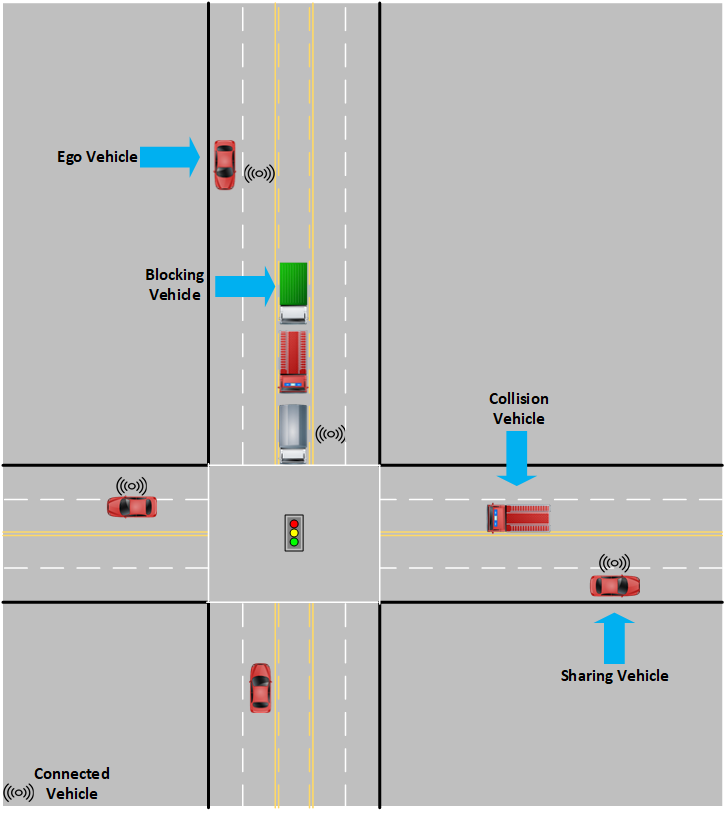}
\caption{Case study scenario: an intersection composed of two types of roads with different vehicles crossing or approaching, designed moving obstacles to create NLOS for both wireless and perception in the turning lane, the traffic light will be ignored by specific vehicles to plan an avoidable collision.}
\label{fig_2}
\end{figure}

\subsection{Case Study}
A generic scenario of mixed vehicles near or approaching a typical intersection is depicted in Figure. 2. The vertical main road has five lanes and the horizontal road has four lanes, with buildings in blocks divided by the roads. Vehicles can be sorted into two main categories: basic vehicles construct the potential collision event, and randomly spawned vehicles represent background traffic. The basic vehicles include the ego vehicle (i.e., vehicle with cooperative perception-based autonomous driving and V2X capability), the collision vehicle ignoring the traffic lights (i.e., vehicle with fine-tuned speed setting to collide with the ego vehicle), the lidar sensor sharing node (i.e., solid perception of the collision vehicle) and the blocking vehicles (i.e., vehicles with significant heights to cause NLOS issues for both the perception system of the ego vehicle and wireless transmission between ego vehicle and sharing node). One randomly spawned vehicle is set to spawn per {\bf{N}} m per lane on average (i.e., {\bf{1000/N}} vehicles per kilometers for each lane), and all randomly spawned vehicles are V2X enabled to be relay node candidates. Such intersection red light violation scenario is chosen not only for its typicality as pre-crash typology from the US National Highway Traffic Safety Administration (NHTSA) but also because it suffers from NLOS-influenced transmission. Along with buildings as static obstacles, all vehicles will be considered as mobile obstacles. All obstacles can cause NLOS conditions and other propagation issues that affect the sensing capability of sensors and wireless transmission of V2X modules. Our goal is to analyze the nonlinear relation between the cooperative perception-based autonomous driving success rate (e.g., completing routes without collision) and V2X communication quality. We therefore emphasize the packet error rate (PER) {\bf{R}} which directly influences the end-to-end performance collision rate ({\bf{CR}}). The two main competitors of V2X technologies, Dedicated Short Range Communications (DSRC) sponsored by the Institute of Electrical and Electronics Engineers (IEEE) and Cellular V2X (C-V2X) endorsed by the 3rd Generation Partnership Project (3GPP), both have pros and cons. To concentrate on studying connection availability and simplicity of simulation development, DSRC is chosen as the underlying vehicular communication technology since our focus is on the impact of NLOS issues on transmission rather than communication efficiency, such as the precise radio resources allocation by eNodeB/gNodeB or the increased data rate through more advanced modulation and coding. 

\subsection{Obstacle Shadowing}
For millimeter wave (mmWave) transmission, 24.25 GHz-52.6 GHz, i.e., Frequency Range 2 (FR2), the beam-type communication is absent as long as NLOS exists, even for Uu type of communication by base stations one block away forming NLOS \cite{ref22, ref23, ref24}, let alone the much less powerful UE-to-UE communication. Frequency bands approved for V2X short-range communication are all sub-6G, mainly for some first-stage basic services and some demos/try-outs of second-stage application scenarios. Sub-6G frequency bands are less susceptible to NLOS interference, but large-scale test results of real roads in China are still not satisfying \cite{ref10}, being prone to struggling with trees and mounds. The signal attenuation effects from static obstacles are not difficult to predict, as buildings, which contribute the most to weakening transmissions, are precisely located on high-definition (HD) maps. The ego vehicle can calculate or predict the obstruction loss along each direction based on the planned route and the HD map. Taking the empirical model \cite{ref25} for example, we can conclude that 9.6 dB signal loss per wall penetrated. On the other hand, moving obstacles pose a height threshold determination issue, and the relative heights of the obstacles and the antennas of V2X enabled vehicle actors are as follows:
\begin{equation}
 \text{$\upsilon$} = {h}\sqrt{\frac {1}{\lambda}(\frac {1}{d_{1}} + \frac {1}{d_{2}})}.
\end{equation}
where \emph{h} stands for the relative height between the obstacle peak to the direct line connecting Tx UE and Rx UE, for DSRC/C-V2X D2D, the wavelength $\lambda$ is approximately 5 centimeters, and \emph{d$_1$}, \emph{d$_2$} are the distances splitted by the obstacle peak. We primarily consider the situation when \emph{h} is larger than 0, the power loss is calculated by the ITU-R recommended equation:
\begin{equation}
 \text{$L_{v-shadow}$} = 6.9 + 20\log_{10}(\sqrt{(\upsilon - 0.1)^{2} + 1} + \upsilon - 0.1).
\end{equation}
The relative height and closer distances dominate the signal loss, with trucks up to approximately 4 m and buses up to 3 m tall. The normal antenna height for a sedan or hatchback is approximately 1.5 m to 1.8 m, and an SUV/MPV may have an extra 0.3 m in height. The shadow effect by calculation can be up to 15-25 dB per large vehicle obstacle, which corresponds to real road tests \cite{ref26, ref27}.

\subsection{Cooperative Perception-based driving control}
We adopted the design of COOPERNAUT \cite{ref9} as the backbone for cooperative perception-based autonomous driving. The autonomous driving policy $\pi$ for ego vehicle is derived based on the observation \emph{O$_t$} received at time \emph{t}. The observation \emph{O$_t$} could be a combination of \emph{O$_{t\_ego}$} and \emph{O$_{t\_i}$} as \emph{i} represents the index of neighboring \emph{N$_t$} vehicles, where \emph{i} $\in$ \{1,...,\emph{N$_t$}\}. $\pi$(\emph{a$_t$}$\vert$\emph{O$_{t\_ego}$}, O$_{t\_1}$, . . . , O$_{t\_N_t}$). \emph{a$_t$} is a set of control decisions on petrol, brake and steering. A trained autonomous driving agent with a hypothetically constant 90\% more data packet reception rate for sensor fusion can easily achieve a 0\%-1\% collision rate in a constrained scenario  \cite{ref9,ref33}. However, the dynamically impaired QoS of V2X wireless communication will lead to nonlinear driving performance degradation, which was overlooked by previous papers \cite{ref9,ref33}, and therefore we more comprehensively consider the entire system.

When combined with the designed traffic scenario and obstacle shadow effects introduced in the previous two subsections, the blocking vehicles do not affect the transmission path between the ego vehicle and the sharing node at the beginning to enable sensor fusion triggering. However, they do obstruct most of the packets when the ego vehicle is approaching the intersection, severely degrading the performance of cooperative perception-based autonomous driving due to the lack of essential shared perception of the collision vehicle.

\subsection{Problem analysis}
In most regions with a large potential market for massive V2X deployment, the permitted V2X frequency bands are limited. These bands would enable multiple services, such as cooperative lane changing, platooning, and vulnerable road user (VRU) notifications by transmitting standard V2X messages (i.e., BSM, RSM, CAM). Even when deploying state-of-the-art sensor fusion algorithms, the required bandwidth is still at 2-10 Mbps, corresponding to a 3-10 MHz bandwidth depending on the specific V2X technologies and modulation schemes. This means that only a few parallel sensor fusion processes can be supported even with ideal channel assignment. In the future, more frequency bandwidth might be agreed globally, but bandwidth for V2X will still be regarded as a precious resource for sensor fusion services. Thus, the initial step for sensor fusion/cooperative perception application is validating the necessity for triggering such service with minimum computation and communication costs. We also investigate the impact of the data compression ratio on collision, as the backbone of the lidar point encoder \cite{ref9} is based on Transformer and down-sampling, which could be adjusted to form an alternative shared feature size for later fusion.

In addition to the bandwidth limitation, we emphasize NLOS transmission, as this is the most critical issue identified in our field test in Wuhan \cite{ref10}. With such impaired QoS of V2X communication, the effective communication range will fall from 100 m to 50 m with only one large vehicle shadowing \cite{ref23, ref24}, and corner cases of truck platoons may completely obstruct the transmitted signal as in our designed traffic scenario. Effective and continuous shared perception of surrounding objects will be impaired by such communication conditions, thus leading to unexpected accidents. To address the issues mentioned above, we propose APMM and MoHeD to trigger sensor fusion and reliable relay finding, respectively, co-determining an optimized V2X communication policy.

\section{Proposed Solution}
\subsection{Sensor Fusion Match \& Trigger}
To achieve a quick and accurate sensor fusion match process for varying scenarios, we introduce an abstract perception matrix (APM) for matching the perception demand of the ego vehicle to the ideal perception provider, which is inspired by the pillar processing of point cloud data \cite{ref34} and bird's eye view map segmentation. Based on the point cloud data collected by one or multiple sensors mounted on the sensor sharing enabled vehicle, raw data points are projected onto a predefined 2D grid after filtering and pruning, and the index value of each grid is assigned by counting the number of points projected. Then, the raw data perception area is down-sampled into a matrix (\emph{M$_{i}$}) representing the concentration of \emph{m$\times$n} grid formed rectangle, and each grid stands for a \emph{k$\times$k} meter sized area as the resolution of the APM with an abstract perception index value. Such matrix data could be broadcasted by vehicle or RSU periodically with a small payload (e.g., 1600 bytes for a \emph{20$\times$20} matrix), along with the absolute coordination (e.g., GNSS location) of the source UE vehicle and its center location corresponding to the matrix. For the fusion consumer vehicle (i.e., ego vehicle in our designed scenario), with its own APM (\emph{M$_{ego}$}) generated, use filter matrix with varying window sizes (\emph{m$_{filter\_i}$}) to locate the area(s) below the perception info threshold (\emph{T$_{1}$}), which could be regraded as blind zone representation(s) with corresponding center and range (\emph{x$_{i}$, y$_{i}$, r$_{i}$}). Blind zone representation(s) on the APM(s) received from surrounding sensor sharing node(s) are mapped by coordinate transformation matrix. The overlapping areas of blind zone representation and the APM are in the form of another sub-matrix (\emph{m$_{benefits}$}), which is used to calculate the perception benefit value by using the sum of the abstract perception index multiplied by the area(s) of overlapping. The benefit outcomes from each sensor sharing node are compared to the predefined thresholds, and the sensor sharing node(s) for initiating data transmission are validated for actual cooperative perception. Unlike the traditional text-based detected vehicle list by each sensor sharing node in the form of cooperative awareness messages (CAMs) or roadside safety messages (RSMs), our APMM method enables low computation and a standardized message payload-friendly match process to determine whether the sensor fusion process needs to be triggered.

\begin{figure}[h]
\centering
\includegraphics[width=3.4in]{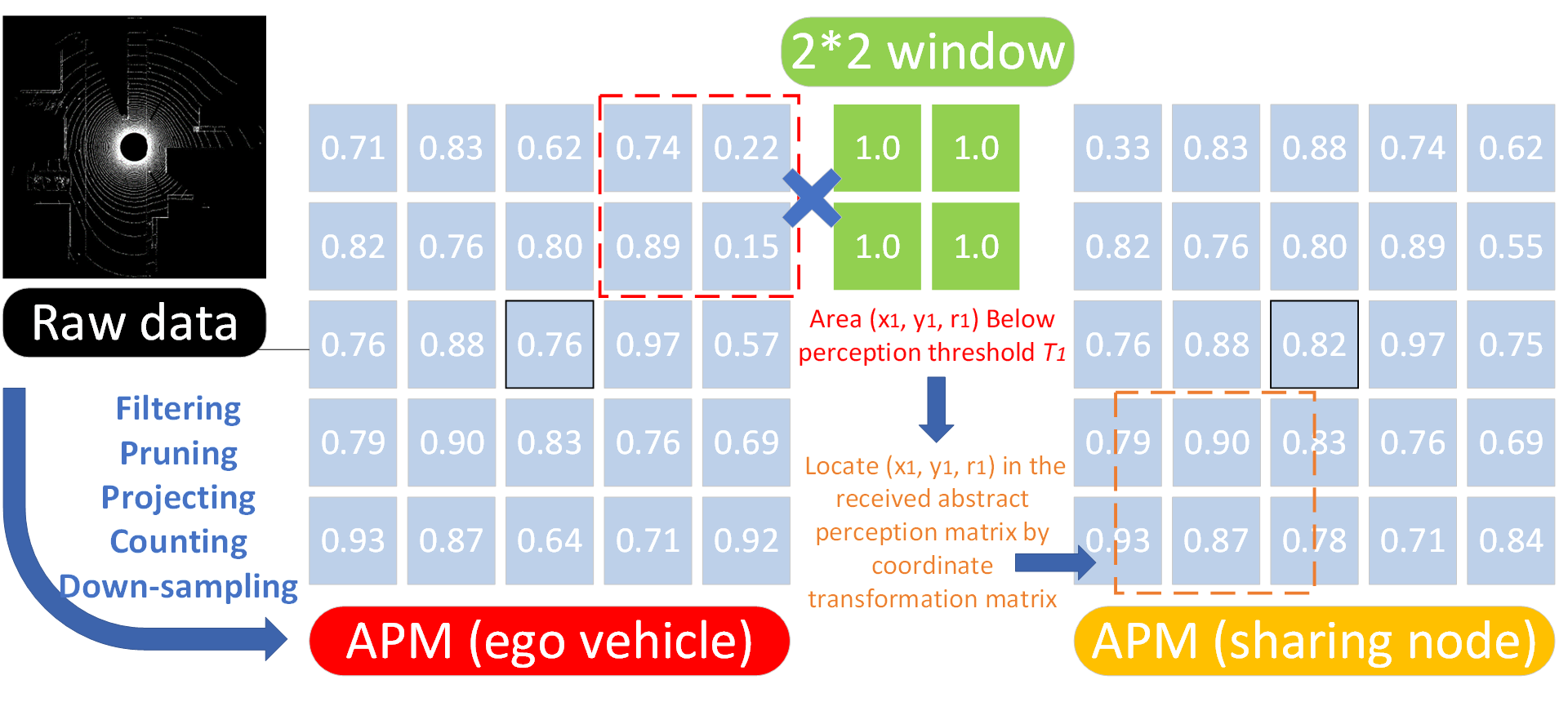}
\caption{Workflow of APMM: process lidar raw data to APM, locate the insufficient perception zone of the ego vehicle (i.e., perception demand of the perception impaired vehicle) by filter windows, and qualify the sensor sharing node candidate by using APM to calculate the sharing perception benefits of overlapping areas.}
\label{fig_3}
\end{figure}

\subsection{Relay Selection}
Once the sensor sharing node is determined, reliable data transmission should be guaranteed. In our designed traffic scenario, the relay node is indispensable because blocking vehicles obstruct the signal from the target sharing node when the ego vehicle approaches the intersection. We introduce the MoHeD method to select appropriate relay node(s). Based on the sensing capability of the ego vehicle, lidar-based object detection and the tracking algorithm could recognize surrounding vehicles. Those vehicles could be targeted as mobile obstacles with effective blocking heights and corresponding mobility information in the mobility and height matrix (i.e., an additional matrix attached to the APM). By exchanging basic information (e.g., utilizing standardized V2X Basic Safety Messages), the ego vehicle could be provided with the mobility and physical size information of each communication node (i.e., the ego vehicle, the relay node, the sensor sharing node). The APMs received from the other vehicle actors introduced in the previous subsection could also be utilized if the mobility and height matrix of vehicle-type obstacles are added from different sides to enrich the obstacle objects information by adding another layer to reveal the representative height and mobility statistics information of each grid. Upon receiving the APM(s) with the mobility and height matrix from the sensor sharing node or relay node candidates as a trigger, the ego vehicle actor will prioritize the relay node candidate(s) (e.g., vehicle or RSU) in the following order within a predefined time window (in our case, 2000 milliseconds):
\begin{enumerate}
\item{Locate the sensor sharing node and the relay node candidates on the mobility and height matrix of the ego vehicle with the basic height and mobility information of each node. This information contains the existing mobility and height information of obstacles sensed by the ego vehicle}
\item{If available, combine the additional mobility and height matrix from the perspective of other nodes with the sensing result of the ego vehicle to generate more comprehensive background info}
\item{Filter out grid(s) representing a valid obstacle in a sub-matrix formed by the index value of communication nodes, as illustrated in Figure 4.}
\item{For each relay node candidate, assign the NLOS risk value by checking whether the obstacle object is located between two nodes as depicted in Figure 4, using the Euclidean distance of the velocities of the communication nodes for NLOS duration approximation}
\item{Choose a direct link or choose the relay node with less NLOS risk by comparing the NLOS risk values calculated in the previous step}
\end{enumerate}
The proposed MoHeD method proposed originates from the intuitive thinking of checking the propagation line whether intersects any objects’ physical outline, one step further into stable and long-lasting LOS communication channel, the tendency to break or to maintain the NLOS path is measured by the mobility characteristics' similarity (\emph{S$_{i}$}) of communication nodes and obstacle objects by Euclidean distance as:
\begin{equation}
 \text{$S_{i}$} = \frac {1}{|v_{node} - v_{obstacle\_i}|} + \frac {1}{|v_{ego} - v_{obstacle\_i}|}.
\end{equation}
then combine all NLOS risk values related to the specific communication node to reflect the NLOS risk value (\emph{V$_{NLOS}$}) as:
\begin{equation}
 \text{$V_{NLOS}$} = \sum _{i=1}^{n} L_{v-shadow\_i}S_{i}.
\end{equation}

\begin{figure}[h]
\centering
\includegraphics[width=3.4in]{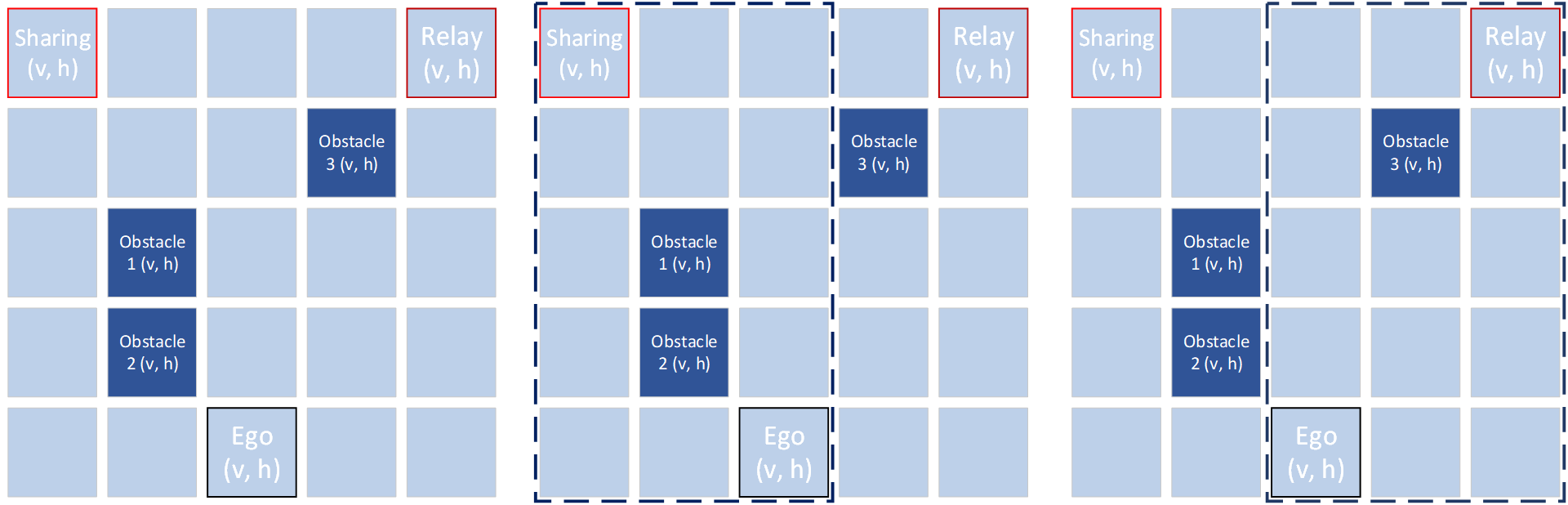}
\caption{Based on the combined mobility and height matrix, schematic diagram of submatrix determination by index value range of ego vehicle and other communication nodes to reduce the search range of NLOS obstacles.}
\label{fig_4}
\end{figure}

\begin{figure}[h]
\centering
\includegraphics[width=1.8in]{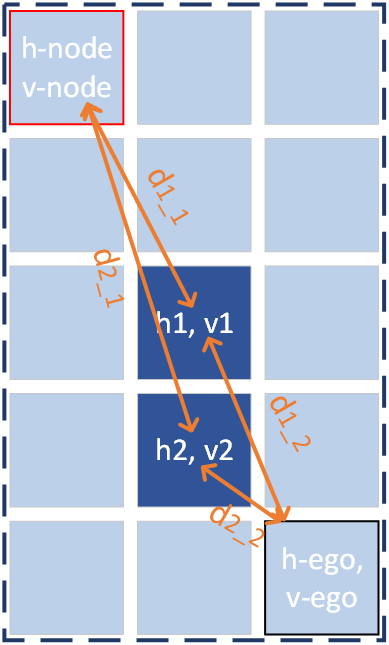}
\caption{Schematic diagram of NLOS effect influence factor derivation corresponding to one specific communication node candidate, taking both mobility and height factor of obstacles, the ego vehicle, and the target communication node into consideration.}
\label{fig_5}
\end{figure}

\begin{figure*}[t]
\centering
\includegraphics[width=6.5in]{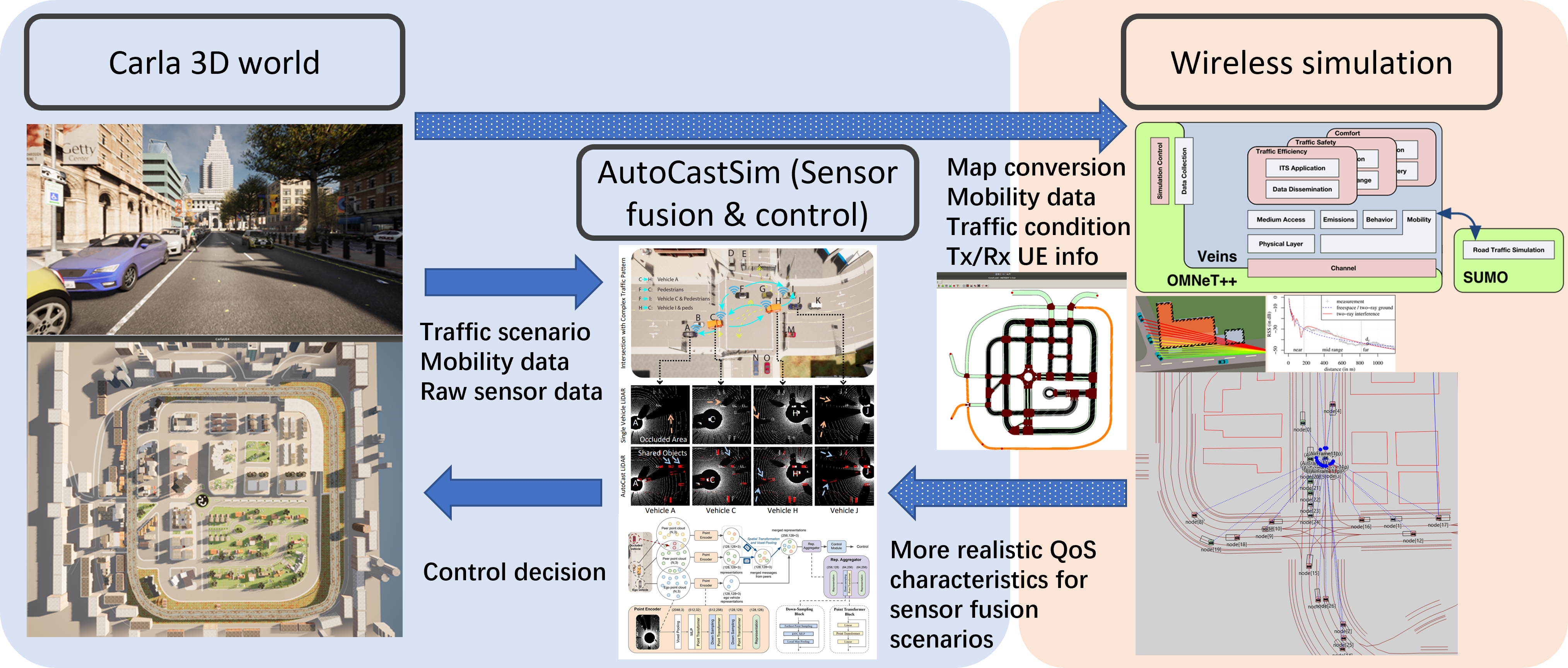}
\caption{Simulation framework illustration: bridging mobility, sensing and communication.}
\label{fig_6}
\end{figure*}

\section{Simulation \& Results}
\subsection{Simulation Framework and Settings}
As depicted in Figure.6, the overall simulation framework could be divided into three parts: 
\begin{enumerate}
\item{The CARLA \cite{ref31} 3D world for traffic scenario construction, vehicle spawning and control, sensor deployment, data extraction, etc., provides fundamental materials for sensor sharing, wireless communication simulation and autonomous driving purposes. We deploy the intersection collision traffic scenario in the CARLA world and extract all map data, physical characteristics of vehicles and buildings, traffic mobility data, sensor data, etc., required for sensor fusion-based autonomous driving and wireless simulation. The wireless QoS-influenced sensor fusion-based autonomous driving control decisions on the CARLA instanced ego vehicle are utilized in a real-time manner.}
\item{The actual sensor fusion and control module executed by the AutoCastSim from COOPERNAUT \cite{ref9}, with a lightweight end-to-end cooperative perception-based autonomous driving model deployed to make the control decision of the ego vehicle based on fusion results from accessible sensor data. We develop a V2X wireless QoS characteristics function, taking input from the wireless simulation section, to influence the data packet transmission for the sensor fusion process.}
\item{The wireless simulation section, where V2X wireless communication is simulated considering the mobility data of the vehicles, the physical sizes of the vehicles and the buildings, constructs a more realistic V2X channel model for wireless QoS characteristics analysis. We develop the wireless simulation section based on Veins \cite{ref30}, which is built on OMNeT++ \cite{ref29} and SUMO \cite{ref28}. Existing models like SimpleObstacleShadowing for building and VehicleObstacleShadowing \cite{ref25} are loaded for obstacles shadowing effects on signal propagation in addition to the basic simple path loss of signal transmission.}
\end{enumerate}
The key to bridging the wireless simulation section and the other two components are SUMO (Simulation of Urban MObility) \cite{ref28}, which provides TraCI (Traffic Control Interface) to interact with other simulation platforms like CARLA \cite{ref31} and OMNeT++ \cite{ref29}, enabling co-simulation for mobility-based process analysis purposes. We mainly develop modules for connecting CARLA to wireless simulation and wireless simulation to AutoCastSim [4], extracting and converting traffic mobility data and wireless QoS characteristics data to accommodate the requirements of different simulation section. With our unique comprehensive simulation framework, the lidar sensor data fusion phase could be influenced by the wireless condition aligned with the physical traffic scenario rather than some averaged wireless QoS characteristics measured from irreverent road tests.
The setting of the vehicular wireless network simulation is illustrated in Table 1. All the listed parameters can be modified in our simulation to simulate the desired deployment condition. In addition, we also enable relay re-selection with a window size of 2000 milliseconds to determine the updated relay node if necessary.
\begin{table}[h]
\caption{Simulation Parameters\label{tab:table1}}
\centering
\begin{tabular}{|c|c|}
\hline
Section & Value\\
\hline
Transmission Power & 400 mW/ 26 dBm\\
\hline
Noise Floor & -98 dBm\\
\hline
Receiver Sensitivity & -94 dBm\\
\hline
Bitrate (maximum) & 6 Mbps\\
\hline
Carrier Frequency & 5.9 GHz\\
\hline
Bandwidth & 10 MHz\\
\hline
Sensor frequency & 10 Hz\\
\hline
Packet size & 200 bytes\\
\hline
\end{tabular}
\end{table}

\subsection{Results \& Analysis}
We use the collision rate of the autonomous driving vehicle (i.e., the ego vehicle in the designed scenario) as the end-to-end performance metric, and the packet error rate (i.e., 1 - packet reception rate) is taken as the main QoS characteristic at current stage. Different levels of compression rate are tested by adjusting the combination of down-sampling and Transformer blocks of the Point Encoder \cite{ref9} before sensor fusion, the collision rate corresponding to a higher compression rate (i.e., 32 times) increased to an unacceptable value even under perfect wireless connection, as depicted in Figure 9. The smaller feature map generated from raw lidar data may not contain sufficient semantic information to reconstruct the unobstructed perception, which leads to catastrophic collision results at an approximately 40\% collision rate. Meanwhile, even acceptable performance achieved with smaller sizes of data (e.g., 3 Mbps/6 Mbps for compression rate at 32/16 times) does not actually contribute to the NLOS situation, as data with or without re-transmission just cannot be delivered due to the high propagation loss. In our designed scenario, the direct link from the ego vehicle to the sensor sharing node is severely obscured by the blocking bus, leading to a PER of 60.78\%. Therefore, the relay is the only way to ensure that the sensor fusion plays its role under such traffic scenarios, counteracting the NLOS issues of wireless transmission for addressing the NLOS issues of perception.

Having a long history of studying D2D relays, 3GPP in Rel-18 has made the newest conclusion on D2D relay selection based on the measured signal strength of the relay discovery solicitation message \cite{ref35}. Signal strength measurements of both D2D direct connections (i.e., End UE to Relay UE) are utilized to determine the most suitable relay node for the one-hop UE-to-UE relay selection procedure. In addition to the 3GPP standard aligned relay selection mechanism, we also choose random relay selection and no relay condition as the baselines for comparison to our proposed mobility-based relay selection method. Based on the wireless performance under different densities of background traffic presented in Figure 7 and the distribution of the packet reception rate on the CDF graph in Figure 8, our mobility-based method outperforms the other two methods for relay selection by at least 25\% enhancement in packet reception rate and has a more concentrated data distribution on the CDF graph. As depicted in Figures 9 and 10, our mobility-based relay selection method is the only one to guarantee a basic functional cooperative perception-based autonomous driving performance, exploiting the sensor fusion benefits for autonomous driving, the packet error rate at 12.02\% brings the collision rate down to less than 1\%. The speed factor of vehicles does affect the actual collision results, but the clear critical point remains the same. One counter-intuitive statistic is that the packet reception rate of the signal-strength-based relay method is worse than that of random relay node selection, this mainly originates from the condition that when a relay node has the in-sum strongest signal strength in both links, it is highly possible for the selected relay node to be close to both the transmitting node and the receiving node while approaching the NLOS area. This leads to an instant best signal strength value that will suffer from the upcoming NLOS areas. Our method overcomes this issue by minimizing the potential intersection of the predicted trajectory of the relay node and the NLOS area and by minimizing the duration of NLOS channel formation by considering the mobility similarity factor. Reflecting this, the number of relay re-selection falls from 3.34 times and 3.65 times for the two bench-marking methods to 2.75 times for our method as shown in Table.II.

\begin{table}[h]
\caption{Performance of relay Node selection methods\label{tab:table1}}
\centering
\begin{tabular}{|c|c|c|}
\hline
Method & Average PRR & Relay Switch times\\
\hline
Mobility-based & 87.98\% & 2.75\\
\hline
Signal-strength-based & 46.59\% & 3.34\\
\hline
Random Selection & 51.48\% & 3.65\\
\hline
Direct Link (no relay) & 39.22\% & 0\\
\hline
\end{tabular}
\end{table}

\begin{figure}[!t]
\centering
\includegraphics[width=3.4in]{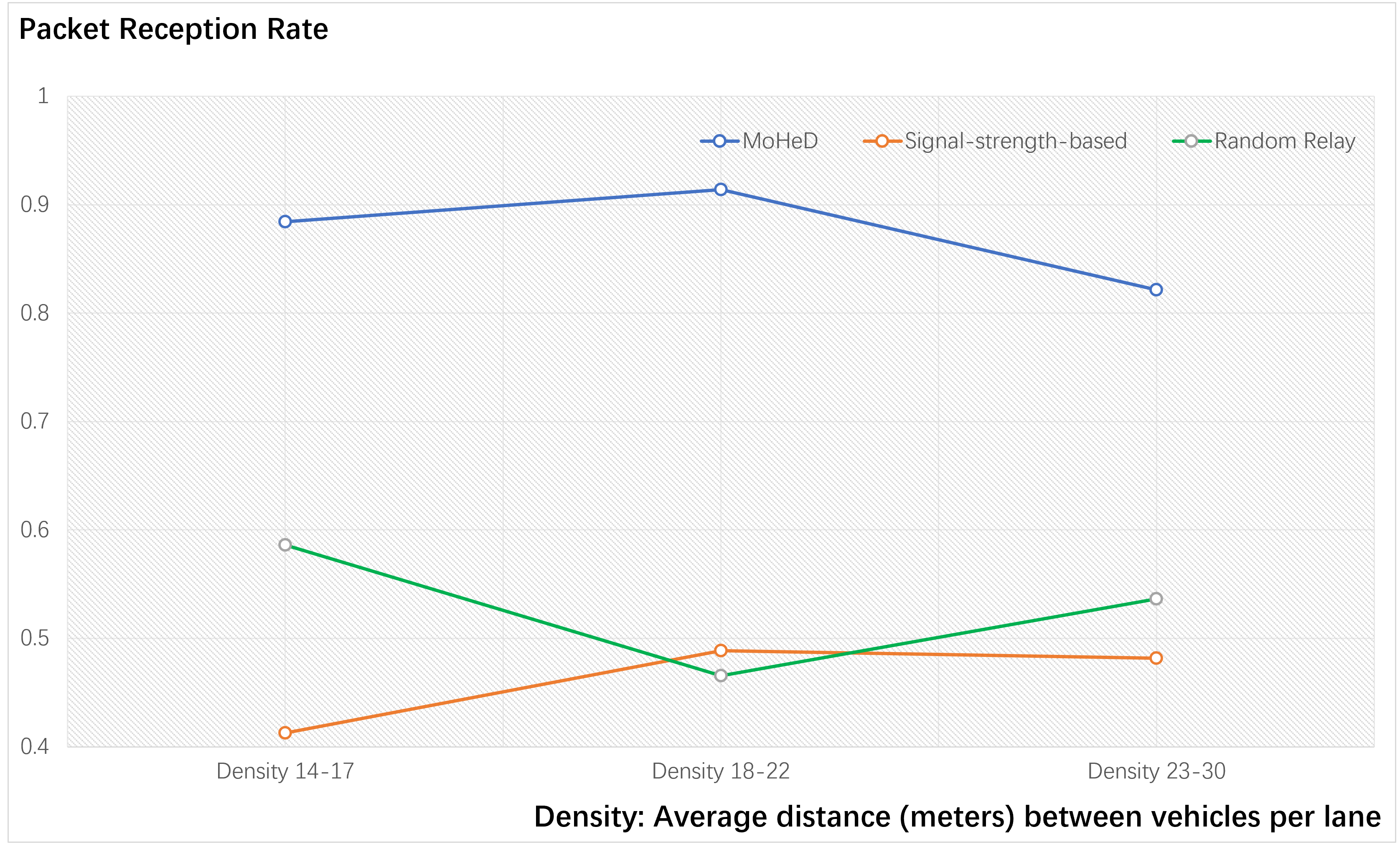}
\caption{Average packet reception rate on different vehicle densities, with ego vehicle's target speed set to 30 km/h and compression rate at 32 times.}
\label{fig_7}
\end{figure}

\begin{figure}[!t]
\centering
\includegraphics[width=3.4in]{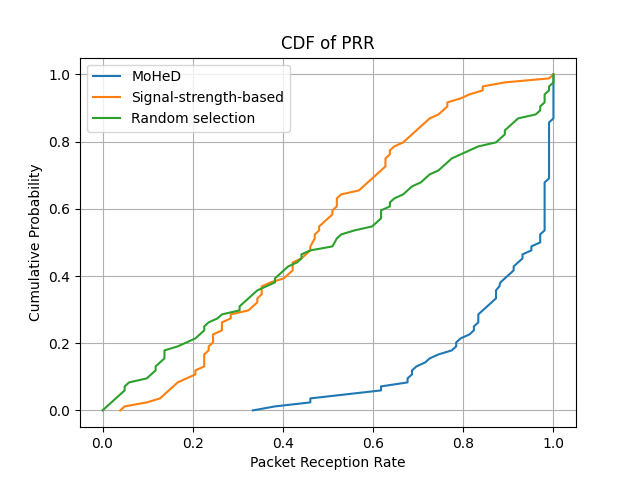}
\caption{CDF of packet reception rate using different relay selection methods, with the target speed of the ego vehicle set to 30 km/h and compression rate at 32 times.}
\label{fig_8}
\end{figure}

\begin{figure}[!t]
\centering
\includegraphics[width=3.4in]{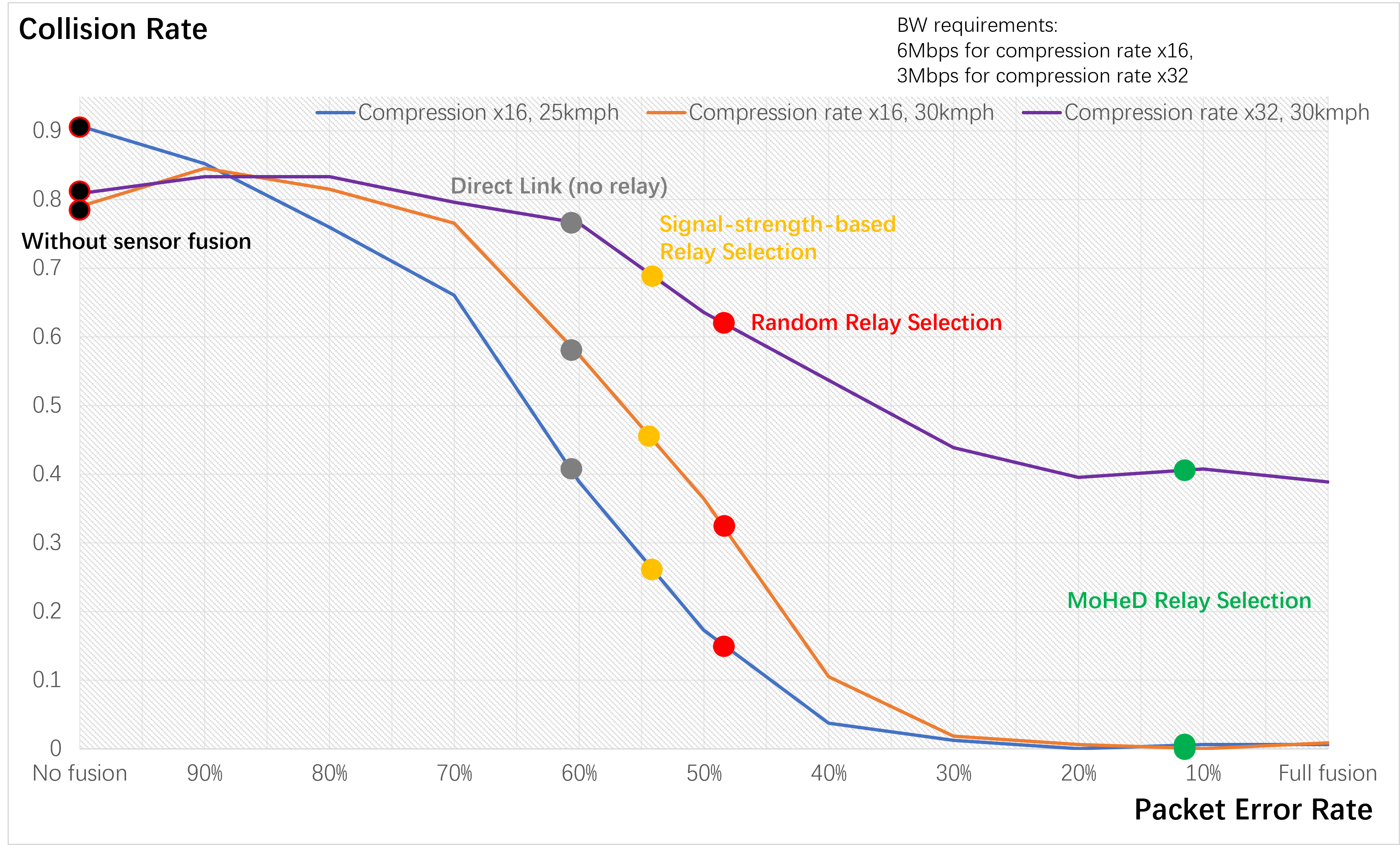}
\caption{Analysis of the collision rate (ego vehicle) and packet error rate: statistics under different compression rates of processed data (compression rate at 16 times and 32 times), different mobility settings (ego vehicle's target speed at 30 km/h and 25 km/h) and different relay selection methods.}
\label{fig_9}
\end{figure}

\section{Conclusion}
In the real world, numerous accidents result from blind zones in the perceptions of human driver, and we tackle the typical NLOS situation for machine perception system and wireless communication as well. Through simulation, we demonstrate that our approach on sensor fusion match and communication relay node selection overshadows the existing methods in both stability and availability, possessing high deployment ability in current standard V2X messages.
In future work, we will further explore the sensor fusion method and AI-based cooperative perception strategy. On the one hand, sensor fusion evolves daily in sensor types and fusion algorithms, and the rising concept of semantic communication seems perfect for sensor fusion data transmission purposes. On the other hand, the matrix processing method proposed in this paper and the various environmental characteristics might serve as a solid foundation for constructing neural networks for next-stage performance upgrades.

\end{document}